\documentstyle[12pt]{article}
\newcommand{\be}{\begin{equation}}
\newcommand{\ee}{\end{equation}}
\newcommand{\ba}{\begin{eqnarray}}
\newcommand{\ea}{\end{eqnarray}}

\newcommand{\bi}{\bibitem}

\begin{document}
\begin{center}
{\bf \Large{LAX  TENSORS, KILLING TENSORS  }}
\end{center}
\begin{center}
{\bf\Large AND GEOMETRIC DUALITY}
\end{center}
\begin{center} {\bf Dumitru Baleanu}\footnote[1]
{E-mail: baleanu@venus.nipne.ro,
dumitru@newton.physics.metu.edu.tr}

 { Middle East Technical University , Physics Department-06531
Ankara,Turkey}

\begin{center}
and
\end{center}

{ Institute of Space Sciences, P.O.BOX, MG-23, R
76900,Magurele-Bucharest, Romania}
\end{center}
\begin{center}
{\bf and}
\end{center}
\begin{center}
 {\bf  Ay{\c s}e (Kalkanl\i) Karasu}\footnote[2]
 {E-mail:akarasu@metu.edu.tr}
 \end{center}
\begin{center}
{ Middle East Technical University , Physics Department-06531,\\
Ankara, Turkey}
\end{center}
\bigskip
\nopagebreak
\begin{abstract}

The solution of the Lax tensor equations in the case
$L_{\alpha\beta\gamma}=-L_{\beta\alpha\gamma}$ was analyzed. The
Lax tensors on the dual metrics were investigated. We classified
all two dimensional metrics having the symmetric Lax tensor
$L_{\alpha\beta\gamma}$. The Lax tensors of the flat space,
Rindler system and its dual were found.
\end{abstract}

\section{Introduction}
Killing tensors are indispensable tools in the quest for exact
solutions in many branches of general relativity as well as
classical mechanics \cite{gib}.
 Killing tensors are important for solving the
equations of motion in particular space-times.The notable example
here is the Kerr metric which admits a second rank Killing tensor
\cite{gib}.Killing tensors give rise to new exact solutions in
perfect fluid Bianchi and Katowski-Sachs cosmologies as well in
inflationary models with a scalar field sources \cite{ross1}.
 Recently  Killing tensors of third rank in $(1+1)$ dimensional
geometry were investigated and classified \cite{ros}.Even more
recently the Killing tensors of order two associated with
orthogonal separable coordinates for the Klein-Gordon equation in
flat 2+1 dimensional space-time were considered as metrics
\cite{franz}. In a geometrical setting ,symmetries are connected
with isometries associated with Killing vectors, and more
generally, Killing tensors on the configuration space of the
system.An example is the motion of a point particle in a space
with isometries ,which is a physicist's way of studying the
geodesic structure of a manifold \cite{hol}.We recall that
$K_{\alpha\beta}$ is a Killing tensor if and only if , for any
geodesic motion of a test particle with a world velocity
$p^{\alpha}$,the scalar $K_{\alpha\beta}p^{\alpha}p^{\beta}$ is a
constant of motion \cite{eisen}.The Jacobi's geometrical model of
dynamical systems with a finite number of degrees of freedom was
investigated by many authors (see for example
Refs.\cite{rauch,rosquist}).The essential conclusion was that :the
paths of the motions of a dynamical system in the configuration
space are identical with the geodesics of the Riemannian manifold
obtained by providing the configuration space with the metric
given by
\be
ds^2=g_{ij}dq_idq_j =2(E-V)a_{ij}dq_idq_j. \ee We mention here
that $T={1\over 2}a_{ij}{\dot q_{i}}{\dot q_{j}}$(the dot
signifying derivation with respect to time and $a_{ij}$ are
functions of the q's),V is a function of the q's only and $T+V=E$.
 In \cite{rosg} it was pointed out that a single  Lax tensor may generate an
infinite number of tensors of varying ranks.It is very well known
that the most general constant on a geodesic motion is of the form
\be
K= K_{0} +\chi_{\mu}p^{\mu} +K_{\mu\nu}p^{\mu}p^{\nu} +
K_{\mu\nu\lambda}p^{\mu}p^{\nu}p^{\lambda}+ \cdots , \ee where
$K_{0}$ is a constant of motion on the geodesic , $\chi_{\mu}$ is
a Killing vector  and $K_{\mu_{1}\cdots\mu_{n}}$ is a Killing
tensor of order n.
 The important point is
that if we are using Jacobi's geometrical model a natural way to
produce Killing tensors is to consider the elements of the Lax
matrix $L_{\alpha\beta}$ \cite{lax,perelomov} as
$L_{\alpha\beta}=L_{\alpha\beta}^{\gamma}p_{\gamma}
+C_{\alpha\beta}$ \cite{karlo}.Here $C_{\alpha\beta}$ is a matrix
having the elements satisfying the following relations
$tr(C_{\alpha\beta})=K_{0}$,$L^{\alpha}_{\alpha\beta}=\chi_{\beta}$,
$K_{\alpha\beta}=L^{\mu}_{\nu\alpha}L^{\nu}_{\mu\beta}$ and so on.
 Open three dimensional Toda's case  analyzed  in \cite{rosg}
is a special and very interesting case because the Lax tensor
generates a Killing tensor of order two which is equal to the
metric tensor.We know that in this case Killing tensor of order
two is called trivial (see for more details \cite{kramer}).
 Recently the
geometric duality between a metric $g^{\mu\nu}$ and its
non-degenerate Killing tensor $K^{\mu\nu}$ and the structural
equations of a Killing tensor of order two were analyzed in
\cite{duh,dubna}.
 An interesting example arises
when the manifold admits Killing-Yano tensors \cite{yano} because
they generate Killing tensors.In addition we know that any
manifold having constant curvature admits Killing-Yano tensors and
then it admits Killing tensors.

For these reasons the Lax tensor equations on a given manifold and
its dual  are interesting to investigate.

 The plan of this paper is as
follows:

 In Section 2 the  Lax pair tensors are investigated.In
Section 3 the geometric duality is presented and the Lax tensors
on the dual manifolds are analyzed.In Section 4 the examples are
presented.Section 5 contains our comments and remarks.

\section{Lax pair tensors}

Let us consider a Riemannian or pseudo-Riemannian geometry with
the metric
\be
ds^2=g_{\mu\nu}dq^{\mu}dq^{\nu}.\ee

The geodesic equation can be represented by the Hamiltonian
\be
H={1\over 2}g^{\mu\nu}p_\mu p_\nu, \ee together with the natural
Poisson bracket on the cotangent bundle.The geodesic system has
the form
\be
{\dot q^{\alpha}}=g^{\alpha\mu}p_\mu,{\dot
p_{\alpha}}=\Gamma^{\mu\nu}_{\alpha}p_\mu p_\nu. \ee

The complete integrability of this system can be shown with the
help of a pair of matrices L and A with entries defined on the
phase space and satisfying the Lax pair equation
\cite{lax,perelomov}. \be\label{ecu} {\dot L}=\{L,H\}=[L,A].\ee It
follows from (\ref{ecu}) that the quantities $I_{k}={1\over
k}TrL^{k}$ are all constants of motion. If in addition they
commute with each other $\{I_{k}, I_{j}\}=0$ then it is possible
to integrate the system completely at least in principle. We know
that Lax pair equation is invariant under a transformation of the
form
\be
{\tilde U}=ULU^{-1}, {\tilde A}=UAU^{-1} -{\dot U}U^{-1}.\ee We
see that L transforms as a tensor while A transforms as a
connection. Typically,the Lax matrices are linear in the momenta
and in  the geometric setting that may also be assumed to be
homogeneous. This motivates the introduction of two third rank
geometrical objects ${L^{\alpha}_{\beta}}^{\gamma}$ and
${A^{\alpha}_{\beta}}^{\gamma}$ such that the Lax matrices can be
written as
\be
L=(L^{\alpha}_{\beta})=({L^{\alpha}_{\beta}}^{\mu}p_\mu),
A=(A^{\alpha}_{\beta})=({A^{\alpha}_{\beta}}^{\mu}p_\mu). \ee We
will refer to ${L^{\alpha}_{\beta}}^{\gamma}$ and
${A^{\alpha}_{\beta}}^{\gamma}$ as the Lax tensor and the Lax
connection , respectively.Defining
\be
B=(B_\beta^\alpha)=({B^{\alpha}_{\beta}}^{\mu}p_\mu)=A-\mit\Gamma,
\ee where
$\mit\Gamma=(\mit\Gamma^{\alpha}_{\beta})=({\mit\Gamma^{\alpha}_{\beta}}^{\mu}p_\mu)$
 is the Levi-Civita connection with respect to $g_{\alpha\beta}$, it
then follows that the Lax pair equation takes the covariant form.
 Let us suppose
that a manifold $g_{\mu\nu}$ admits a Lax pair tensors
$L_{\alpha\beta\gamma}, A_{\alpha\beta\gamma}$ in such a way that
\be\label{las}
L_{\alpha\beta\gamma;\delta}+L_{\alpha\beta\delta;\gamma}=
L_{\alpha\mu(\gamma}B^{\mu}_{\vert\beta\vert\delta)}-B_{\alpha\mu(\gamma}L^{\mu}_{\vert\beta\vert\delta)}.
\ee

Here the parenthesis denotes the full symmetrization.
 We know that a Killing tensor of order
n is a symmetric tensor $K_{\mu_{1}\cdots \mu_{n}}$ which
satisfies the following relation:

\begin{equation}\label{kilingul}
D_{(\lambda}K_{\mu_{1}\cdots\mu_{n})}= 0
\end{equation} where $D_{\mu}$ denote covariant derivative.
 and
 Using
(\ref{las}) ,in the case when $L_{\alpha\beta\gamma}$ has only
symmetric part, we find immediately that
$L_{(\alpha\beta\gamma;\delta)}=0$ for
$B^{\alpha}_{\beta\gamma}=0$ . Then  it is a Killing tensor of
order three. Any solution of (\ref{las}) generates an infinite
number of Killing tensors on a given manifold.Of course not all
Killing tensors generated by Lax tensors are independent and some
of them are trivial Killing tensors \cite{kramer}.Another
important observation is that in the case when we have
$g_{\alpha\beta}=L^{\mu}_{\nu\alpha}L^{\nu}_{\mu\beta}$ we can
identify the invariant $I_2$ with the geodesic Hamiltonian.

Let us suppose that the manifold admits a Killing tensors
$K_{\alpha\beta}$ and define a three dimensional tensor as
 \be\label{tens}
L_{\alpha\beta\gamma}=K_{\beta\gamma;\alpha}-K_{\alpha\gamma;\beta}.
\ee We conclude  immediately that it has the symmetries
\be
L_{\alpha\beta\gamma}=L_{[\alpha\beta]\gamma},L_{[\alpha\beta\gamma]}=0,
\ee

where  square brackets denote the anti-symmetrization. After an
appropriate grouping of terms and use of the symmetries of the
Riemann tensor $R_{\alpha\beta\gamma\delta}$  we obtain
\be\label{ltrei}
L_{\alpha\beta(\gamma;\delta)}=-2R_{\alpha\beta\mu(\gamma}
K_{\delta)}^{\mu}
-2K^{\mu}_{[\alpha}R_{\beta]\mu(\gamma\delta)\mu}. \ee

We are interested now to investigate if (\ref{tens}) satisfies
(\ref{las}).In other words our problem is to find a tensor
$B_{\alpha\beta\gamma}$ in such a way that (\ref{las}) is
satisfied.Using (\ref{tens}) and (\ref{las}) we conclude that
$B_{\alpha\beta\gamma}=-B_{\beta\alpha\gamma}$.
 Let
us denote $ V_{\alpha\beta\gamma}$ as\\
$V_{\alpha\beta\gamma}=L_{[\alpha\beta]\gamma}$.Taking into
account (\ref{las}) and (\ref{ltrei}) we find that
\be\label{lpatru} V_{\alpha\mu(\gamma}
B^{\mu}_{\vert\beta\vert\delta)}-V_{\beta\mu(\gamma}
B^{\mu}_{\vert\alpha\vert\delta)}= R_{\alpha\beta\mu(\gamma}
K_{\delta)}^{\mu}
+K^{\mu}_{[\alpha}R_{\beta]\mu(\gamma\delta)\mu}.\ee Solving
(\ref{lpatru}) we can determine $B^{\alpha}_{\beta\gamma}$.

Using definition of a Killing tensor of order two and (\ref{tens})
we get \be\label{asa} K_{\beta\gamma;\alpha}={2\over
3}L_{\alpha(\beta\gamma)}.\ee

Conversely , (\ref{asa})  and the conditions
$L_{\alpha\beta\gamma}=-L_{\beta\alpha\gamma},
L_{[\alpha\beta\gamma]}=0$ imply \\ (\ref{tens}) and that
$K_{\alpha\beta}$ is a Killing tensor.

  Another interesting case is when
$L_{\alpha\beta\gamma}=V_{\alpha\beta\gamma}$ and in addition we
suppose that  $B_{\alpha\beta\gamma}=L_{\alpha\beta\gamma}$.For
this case the Lax equations become \be\label{ert}
V_{\alpha\beta(\gamma;\delta)}=0, \ee and we see that (\ref{ert})
looks like Killing-Yano equations.

\section{Geometric Duality}

  Let  us suppose that the metric $g_{\mu\nu}$ admits
a Killing tensor field $K_{\mu\nu}$.

{}From
  the covariant components $K_{\mu\nu}$ of the Killing tensor one can
construct a constant of motion
$K=\frac{1}{2}K_{\mu\nu}p^{\mu}p^{\nu}$. It can  be verified that
$\{H,K\}=0$. The formal similarity between the constants of motion
H and K , and the symmetrical  nature of the condition implying
the existence of the Killing tensor amount to a reciprocal
relation between two different models:the model with Hamiltonian H
and constant of motion K, and a model with constant of motion H
and Hamiltonian K.The relation between the two models has a
geometrical interpretation: it implies that if $K_{\mu\nu}$ are
the contravariant components of a Killing tensor with respect to
the metric $g_{\mu\nu}$, then $g_{\mu\nu}$ must represent a
Killing tensor with respect to the metric defined by $K_{\mu\nu}$.
 When $K_{\mu\nu}$ has an inverse we interpret it as the
metric of another space and we can define the associated
Riemann-Christoffel connection $\hat\Gamma_{\mu\nu}^{\lambda}$ as
usual through the metric postulate ${\hat
D}_{\lambda}K_{\mu\nu}=0$. Here ${\hat D}$ represents the
covariant derivative with respect to $K_{\mu\nu}$.
This reciprocal relation between the metric structure of pairs of
spaces constitutes a duality relation: performing the operation of
 mapping a Killing tensor to a metric twice leads back to the original
theory.

The relation between connections ${\hat
\Gamma^{\sigma}_{\alpha\beta}}$ and
$\Gamma^{\sigma}_{\alpha\beta}$ is \cite{dubna}

\be\label{cone} {\hat
\Gamma^{\mu}_{\alpha\beta}}=\Gamma^{\mu}_{\alpha\beta}-
K^{\mu\delta}D_{\delta}K_{\alpha\beta}. \ee In the case when the
tensor $B^{\alpha}_{\beta\delta}$ is symmetric in the lower
indices and has the form
\be
B^{\alpha}_{\beta\delta}=
K^{\alpha\omega}D_{\omega}K_{\beta\delta}\ee then (\ref{las})
becomes \be\label{lasi} {\hat L}_{\alpha\beta\gamma;\delta}+{\hat
L}_{\alpha\beta\delta;\gamma}=0. \ee Here comma represents the
covariant derivative in the dual space. We are interested now to
investigate when the original space and its dual admit the same
Lax tensors.

 {\bf Proposition }\\

The manifold  and its  dual have the same Lax tensors iff

\ba\label{laxuri}
&2(K^{\sigma\omega}D_{\omega}K_{\gamma\delta})L_{\alpha\beta\sigma}&
+
(K^{\sigma\omega}D_{\omega}K_{\alpha\delta})L_{\sigma\beta\gamma}
+
(K^{\sigma\omega}D_{\omega}K_{\alpha\gamma})L_{\sigma\beta\delta}
+\cr
&(K^{\sigma\omega}D_{\omega}K_{\beta\delta})L_{\alpha\sigma\gamma}&
+
(K^{\sigma\omega}D_{\omega}K_{\beta\gamma})L_{\alpha\sigma\delta}=0.
\ea

{\bf Proof.}\\ Let us consider $L_{\alpha\beta\gamma}$ the  Lax
tensor satisfies \be\label{lix}
L_{\alpha\beta(\gamma;\delta)}=0
\ee and ${\hat L}_{\alpha\beta\gamma}$ be the dual Lax tensor.
 Using
(\ref{cone})  the  corresponding  dual Lax equations are

\ba\label{laxdual} &D_{\delta}{\hat L}_{\alpha\beta\gamma}&+
D_{\gamma}{\hat L}_{\alpha\beta\delta}+
2(K^{\sigma\omega}D_{\omega}K_{\gamma\delta}){\hat
L}_{\alpha\beta\sigma} +
(K^{\sigma\omega}D_{\omega}K_{\alpha\delta}){\hat
L}_{\sigma\beta\gamma} +\cr
&(K^{\sigma\omega}D_{\omega}K_{\alpha\gamma}){\hat
L}_{\sigma\beta\delta}&+
(K^{\sigma\omega}D_{\omega}K_{\beta\delta}){\hat
L}_{\alpha\sigma\gamma}
 +
(K^{\sigma\omega}D_{\omega}K_{\beta\gamma}){\hat
L}_{\alpha\sigma\delta}=0. \ea
 Let us  suppose that  ${\hat L}_{\alpha\beta\gamma}=L_{\alpha\beta\gamma}$,
 then using (\ref{lix}) and (\ref{laxdual}) we obtain
 (\ref{laxuri})
 Conversely if we suppose that (\ref{laxuri}) holds , then from (\ref{laxdual})
 we  can deduce  immediately that  $L_{\alpha\beta\gamma}={\hat L}_{\alpha\beta\gamma}$.
{\bf q.e.d.}

\section{Examples}
In this section we will present some examples when the equations
(\ref{las}) admit solutions.

{\bf A.}

 Let us consider the n-dimensional Euclidean space and first
 let us investigate
 the Lax equations corresponding to $B^{\alpha}_{\beta\gamma}=0$.Then (\ref{las}) becomes
 \be
 {\partial L_{\alpha\beta\gamma}\over\partial x^{\delta}} +
 {\partial L_{\alpha\beta\delta}\over\partial x^{\gamma}}=0.
 \ee

The solution of this equation has the form
\be
L_{\alpha\beta\gamma}=T_{\alpha\beta\gamma\sigma}x^{\sigma}
+V_{\alpha\beta\gamma}, \ee where $T_{\alpha\beta\gamma\sigma}$
and $V_{\alpha\beta\gamma}$ are constant tensors and in addition
$T_{\alpha\beta\gamma\sigma}=-T_{\alpha\beta\sigma\gamma}$.

{\bf B.}

Let us consider now the two dimensional metric
\be\label{metric}
 ds^2=f(u,v)du^2 +g(u,v)dv^2.\ee
  We are interested to investigate the Lax
tensors when $L_{\alpha\beta\gamma}$ is symmetric and
$A_{\alpha\beta\gamma}=\Gamma_{\alpha\beta\gamma}$. The
non-vanishing Christoffel symbols of (\ref{metric}) are \ba
&\Gamma_{11}^{1}=&{{\partial f\over\partial u}\over 2f},
\Gamma_{11}^{2}={-{\partial f\over\partial v}\over 2g},
\Gamma_{21}^{1}=\Gamma_{12}^{1}={{\partial f\over\partial v}\over
2f},\cr &\Gamma_{22}^{2}=&{-{\partial g\over\partial u}\over
2f},\Gamma_{22}^{2}={{\partial g\over\partial v}\over 2g},
\Gamma_{21}^{1}=\Gamma_{12}^{2}={{\partial g\over\partial u}\over
2g}. \ea
 $L_{\alpha\beta\gamma}$
 has four independent components
$L_{111},L_{112},L_{122},L_{222}$ and the independent Lax
equations are \ba\label{ecua}
&L_{11(1;u)}=0&,L_{11(1;v)}=0,L_{11(2;u)}=0,L_{11(2;v)}=0,\cr
&L_{12(2;u)}=0&,L_{12(2;v)}=0,L_{22(2;u)}=0,L_{22(2;v)}=0. \ea

We found after some calculations that if the scalar curvature of
the manifold corresponding to (\ref{metric}) is 0 then the system
(\ref{ecua}) is integrable.

Let us consider now the Rindler system.
 The Rindler system \cite{hinter} is conventionally denoted by $\tau$ and r
 \be
 t=r\sinh {\tau}, x=r\cos{\tau}, 0 < r < \infty, -\infty < \tau <
 \infty,
\ee with coordinates curves (timelike hyperbolas and spacelike
straight lines) given by
\be
x^2-t^2=r^2, {t\over x}=\tanh{\tau}, \ee the metric
\be
ds^2=r^2d\tau^2 -dr^2, \ee and the associated Killing tensor
\be\label{invers} k^{ik}=\left(\matrix{&1-{c\over r^2}&0\cr
&0&c}\right). \ee Here c is a constant. The non-zero Christoffel
symbols are $\mit\Gamma^{2}_{11}=r$, $\mit\Gamma^{1}_{12}={1\over
r}$. Solving (\ref{ecua})  we found the solution of the Lax
equations having the form \ba &L_{122}&=(C_1 {\rm e^{-\tau}} + C_2
{\rm e^{3\tau}})r, L_{112}=(C_1 {\rm e^{-\tau}} +C_2 {\rm
e^{3\tau}})r^2,\cr
 &L_{111}&=-(3C_1 {\rm e^{-\tau}} - C_2
{\rm e^{3\tau}})r^3, L_{222}=-3C_1 {\rm e^{-\tau}} + C_2 {\rm
e^{3\tau}},
 \ea

where $C_1, C_2$ are constants.

The next step is to find a solution of the form (\ref{tens})
corresponding to the Rindler system. Using (\ref{tens}) and
(\ref{invers}) we found immediately the solution having the form
\be
L_{121}=-{r^3(r^2-3c)^2\over c(r^2-c)^2},L_{122}=0. \ee

Let us consider now the tensor \be\label{susi}
K^{\mu\nu}=g^{\mu\lambda}g^{\nu\delta}K_{\lambda\delta} \ee and a
connection defined as in ({\ref{cone}).
Using (\ref{cone}) and taking into account (\ref{susi})and
(\ref{invers}) we found a new metric  having the non-zero
components \be\label{ndual} {d{\hat s}^{2}}=r^{2}{d\tau}^2 +
{c(r^2-c)^2\over(2r^2 +r^4c-2r^2c^2 +c^3)}{dr}^2. \ee
 The scalar
curvature corresponding to (\ref{ndual}) is
$R={4(r^2+c)\over{c(-r^2 +c)^3}}$ ,then this metric has no
symmetric Lax tensors.
\section{Concluding remarks}

In this paper we investigated the Lax equations on a given
manifold and its dual.
When a manifold admits
a Killing tensor $K_{\mu\nu}$ we constructed
 a tensor $L_{\alpha\beta\gamma}$ as $L_{\alpha\beta\gamma}=K_{\beta\gamma;\alpha}-K_{\alpha\gamma;\beta}$
and found the  conditions when it is a Lax tensor.In this case
$L_{\alpha\beta\gamma}$ is antisymmetric in the first two indices
and $B_{\alpha\beta\gamma}$ should have the same property. If in
addition we suppose that
$B_{\beta\alpha\gamma}=L_{\alpha\beta\gamma}$ we found that
(\ref{las}) has the simple form
$L_{\alpha\beta(\gamma;\delta)}=0$. We found the conditions when
the manifold and its dual have the same Lax tensors.
 For the two dimensional manifolds we found that the symmetric Lax
tensors exist if the scalar curvature is zero. The solution of the
Lax equations for the flat space case , the Rindler system and its
dual manifold were found.

 Finding the Lax tensors on the manifolds which admits
 Killing-Yano tensors is an interesting problem
  and it requires further investigation.

\section{Acknowledgements}

 One of the authors (D.B.) would like to thank Ashok Das for
 helpful discussions.
  He also would like to thank TUBITAK and NATO for
financial support  and METU for the hospitality during his working
stage at the Department of Physics. This work is partially
supported by the Scientific and Technical Research Council of
Turkey.


\begin{thebibliography}{99}

\bi{gib}G.W. Gibbons, R.H.Rietdijk and J.W.van Holten, {\sl
Nucl.Phys.} {\bf B 404}, \\42 (1993).
\bibitem{ross1}K.Rosquist and C.Uggla, {\sl J.Math.Phys.} {\bf 32}, 3412 (1991).
\bibitem{ros} M. Karlovini and K. Rosquist, preprint gr-qc/9807051.
\bibitem{franz}F.Hinterleitner, {\sl Annals of  Physics } {\bf 271},23 (1999).
\bi{hol} R.H.Rietdijk and J.W.van Holten, {\sl J.Geom.Phys.} {\bf
11}, 559 (1993).
\bibitem{eisen}
L.P.Eisenhart, {\sl Riemannian Geometry} (Princeton U.P.,
Princeton, N.J.) (1966).
\bibitem{rauch}
H.E.Rauch, {\sl Geodesics and curvature in differential geometry
in the large}, (Yeshiva University, New York, 1959).
\bibitem{rosquist} K.Rosquist, in {\sl The Seventh Marcel Grossmann Meeting.On
Recent Developmnets in Theoretical and Experimental General
Relativity, Gravitation and Relativistic Field Theories},edited
by.R.T.Jantzen and G.M.Keiser (World Scientific, Singapore), 1997
vol.1, p.379.
\bibitem{rosg} K. Rosquist and M.Goliath, {\sl Gen.Rel.Grav.}
{\bf 30} ,1521 (1998).
\bibitem{lax} P.D.Lax, {\it Comm.Pure.Appl.Math.} {\bf 21}, 467 (1968).
\bibitem{perelomov} A.M.Perelomov , {\sl Integrable systems of classical
mechanics and Lie algebra},  I.Birkhauser, 1990.
\bibitem{karlo}
M.Goliath,M.Karlovini,Kjell Rosquist,solv-int/9810011.
\bibitem{kramer} D.Kramer ,H.Stephani, E.Herlt and M.Mac Callum {\sl Exact Solutions of Einstein
Field Equations},Cambridge University Press, 1980.
\bibitem{duh} R.H.Rietdjik, J.W. van Holten  {\sl Nucl.Phys.} {\bf
  B 472 },472 (1996).
\bibitem{dubna}D.Baleanu and S.Codoban , {\sl Gen. Rel. and Grav.} {\bf 31},497 (1999)
,\\
 D.Baleanu, Preprint JINR E5-98-188,\\
D.Baleanu,in {\sl Proceedings of XIth Intl.Conf.Problems of
Quantum Field Theory , Dubna, Russia, 13-17 July, p.66 (Publ.Dubna
1999, Ed. B.M.Barbashov, G.V.Efimov and A.V.Efremov)},\\
D.Baleanu, to be published in {\sl Supplemento ai Rendiconti del
Circolo Matematico di Palermo}.
\bibitem{yano} K.Yano, {\sl Ann. Math.}, {\bf 55}, 328 (1952).
\bibitem{hinter}
 F.Hinterleitner, {\sl Acta. Phys.Slovaca} {\bf 47} , 157 (1997).

\end{thebibliography}
\end{document}